\documentclass[twocolumn,aps,prx,amsmath,amssymb,color,longbibliography,superscriptaddress]{revtex4-2}
\usepackage{stmaryrd}
\usepackage{amsmath}
\usepackage{amssymb}
\usepackage{graphicx}
\usepackage{dcolumn}
\usepackage{bm}
\usepackage{tabularx}
\usepackage{color}

\begin{document}
\begin{titlepage}
\title{Small polaron formation by electron-electron interactions}
\author{Yunfan Liang}
\thanks{These authors contributed equally to this work}
\affiliation{Department of Physics, Applied Physics and Astronomy, Rensselaer Polytechnic Institute, Troy, NY, 12180, USA}
\author{Min Cai}
\thanks{These authors contributed equally to this work}
\affiliation{School of Physics and Wuhan High Field Magnetic Field Center, Huazhong University of Science and Technology, Wuhan, 430074, China}
\author{Lang Peng}
\affiliation{School of Physics and Wuhan High Field Magnetic Field Center, Huazhong University of Science and Technology, Wuhan, 430074, China}
\author{Zeyu Jiang}
\email{jiangz10@rpi.edu}
\affiliation{Department of Physics, Applied Physics and Astronomy, Rensselaer Polytechnic Institute, Troy, NY, 12180, USA}
\author{Damien West}
\affiliation{Department of Physics, Applied Physics and Astronomy, Rensselaer Polytechnic Institute, Troy, NY, 12180, USA}
\author{Ying-Shuang Fu}
\email{yfu@hust.edu.cn}
\affiliation{School of Physics and Wuhan High Field Magnetic Field Center, Huazhong University of Science and Technology, Wuhan, 430074, China}
\author{Shengbai Zhang}
\affiliation{Department of Physics, Applied Physics and Astronomy, Rensselaer Polytechnic Institute, Troy, NY, 12180, USA}
\date{\today}

\begin{abstract}
In a solid, electrons can be scattered both by phonons and other electrons. First proposed by Landau, scattering by phonons can lead to a composite entity called a “polaron”, in which a lattice distortion traps an itinerant electron (or hole) such that the distortion and carrier move in unison as a single particle with larger effective mass. While this is the traditional view of polarons, the rise of 2D systems, especially strongly correlated ones, open the prospect of electron scattering taking on a larger role in spontaneous carrier localization for such material systems. Here, we show that in transition metal halides, such electron-electron interactions can lead to polaron formation even in the absence of lattice distortion. This suggests an alternative direction for polaron formation, transport, and control in solids. This new mechanism of polaron formation is confirmed by first-principles calculation of 2D transition metal halides, CrI$_2$, CoCl$_2$ and CoBr$_2$. These theoretical predictions are supported by scanning tunneling microscopy/spectroscopy measurements of polarons in CrI$_2$.
\end{abstract}

\maketitle
\draft
\vspace{2mm}
\end{titlepage}
Understanding the polaronic characteristics of doped strongly-correlated systems are of both fundamental and practical importance in condensed matter physics. While polaronic effects are typically significant in ionic semiconductors due to the strong electron-phonon coupling and weak dielectric screening\cite{emin2013polarons,giustino2017electron}, they are increasingly becoming associated with strongly-correlated systems. For instance, high temperature superconductivity in p-doped cuprates is widely explained by the electronic correlation of preformed hole polarons in the CuO$_2$ sublayer\cite{morgan2009small,lee2006doping}. In recent years, polaronic physics in two-dimensional (2D) strongly-correlated systems has attracted considerable attention. Compared with their three-dimensional (3D) counterparts, 2D systems have a smaller dispersion that is expected to increase many-body effects\cite{kravchenko2017strongly}. Furthermore, the ease of carrier injection and the accessibility of such states with surface sensitive techniques, such as scanning tunneling microscopy (STM), make 2D systems a promising platform to substantially further the understanding of polarons, either through direct imaging or through observation of their band dispersion\cite{mao2020interfacial,kang2018holstein}.

The properties of polarons sensitively depend on the nature of their interactions. The so-called large polaron corresponds that the propagation of carrier is dressed by a cloud of virtual phonon with polaron radius much larger than lattice constant\cite{Frohlich1,Frohlich2}. In contrast, a small polaron is usually understood as the spontaneous localization of carrier due to lattice distortions which break the translational symmetry of crystal\cite{Holstein1,Holstein2}. While such a distortion increases the energy of the underlying lattice, it can trap a carrier and lead to the formation of a localized electronic state in the bandgap, lowering the overall system energy. In other words, the conventional mechanism of polaron formation, no matter being large or small, is through electron-phonon coupling where the bare charge is dressed with a lattice distortion, leading to an increase of carrier effective mass or a spontaneous localization. However, from the viewpoint of modern polarization theory\cite{resta2007theory}, the modulation of charge density in solids can be, in principle, a purely electronic effect where a spontaneous ordering of electrons breaks the crystal symmetry, without any atomic displacement. Such a possibility is hinted at by the study of electronic dynamics in the iron-based superconductor FeAs, where the doped electron on Fe atom is dressed by the polarization of surrounding As $4p$ states\cite{berciu2009electronic}, resulting in a reasonably increased effective mass. This raises the question of whether a small polaron can be localized, via a purely electronic mechanism, in solids in the absence of any atomic distortion. Such a mechanism would constitute a fundamentally new type of polaron, here dubbed an \emph{electronic polaron}, in which the electron-phonon coupling needs not play any active role. Instead, electron-electron interaction becomes the dominant in polaron formation, with lattice relaxation only being a response to the electronically-localized carrier.

In this work, we propose a theory of the formation of electronic polarons, in which the underlying physics is due to electron-electron interactions. The doped electron gets self-trapped into polaronic state by polarizing the local valence states of surrounding atoms and generating an electronic potential well on the polaron site. In such systems, the kinetic energy increased by localization is mainly balanced by the electronic response of valence states. Through first-principles calculations, we find that 2D strongly-correlated transition metal halides, CrI$_2$, CoCl$_2$ and CoBr$_2$, are ideal candidates to host electronic polarons. The trapping well of polaron can be as large as $-17$ eV for CrI$_2$, of which only $-5$ eV is contributed by atomic relaxation. Taking CrI$_2$ as a prototypical example, we present the details of the electronic structure, which are supported by our scanning tunneling spectroscopy (STS) and microscopy (STM) experiments on high-quality CrI$_2$ single layer grown via molecular beam epitaxy.

Density functional theory (DFT) calculations were performed by VASP code\cite{kresse1996efficient} with projector augmented wave (PAW) potentials. Hybrid functional HSE06 \cite{heyd2003hybrid,heyd2006erratum} was employed in all the calculations, in order to provide more physical treatment of many-body effect\cite{tran2006hybrid,bilc2008hybrid,zhuang2016strong,hellgren2017critical}.  Unless otherwise stated, we used a supercell of size $5\times5\times1$ to simulate the polaron. The hybrid mixing parameter $\alpha$ for CrI$_2$ is chosen to be 0.4 to fit the experimental gap.We also carried out DFT+U\cite{dudarev1998electron} calculations with the U values determined by fitting the experimental gap.

\begin{figure}[tbp]
\includegraphics[width=1.00\columnwidth]{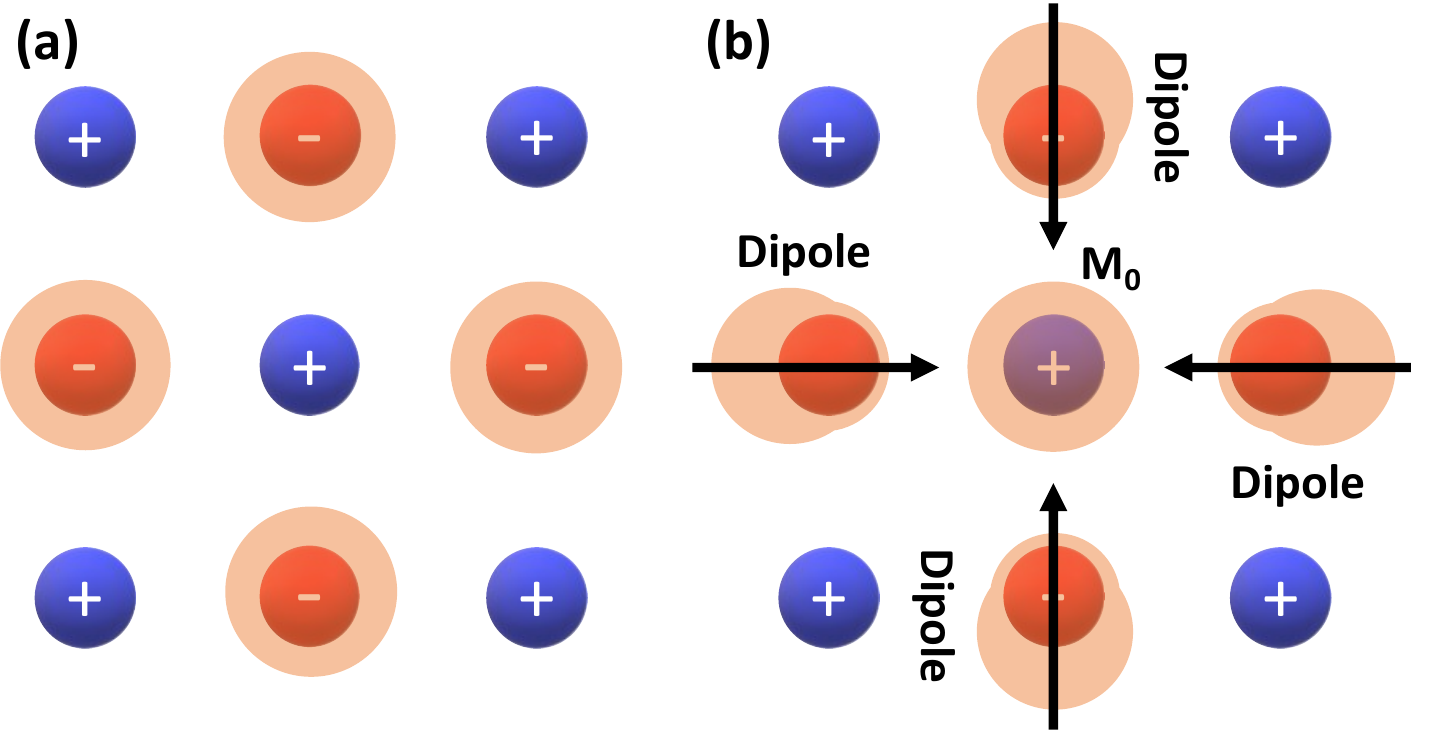}
\caption{\label{fig:fig1} (Color online) (a) The sketch of a 2D system with valence electrons distributed center-symmetrically on the anions. The cation (anion) is colored in blue (red) and the orange clouds denote the distribution of valence electrons. (b) Charge distribution of the electronic polaron state (on the central cation) and the polarized valence electrons (on the anions) under the Coulomb repulsion of electronic polaron. The arrows indicate the dipoles formed by such polarization.}
\end{figure}

Let's consider a single electron doping of 2D model lattice with cation (anion) labeled as + ($-$) in Fig. 1. If the doped electron locates on cation site M$_{0}$, due to the repulsive Coulomb interaction, the valence electrons of surrounding anions will move away from the M$_0$, even with the atoms fixed. The repulsion breaks the symmetric distribution of valence electrons on anions and induces electronic dipoles pointing to the M$_{0}$ site. This will produce an electronic potential well centered on M$_{0}$ which lowers the energy of the doped electron. However, such localization of the doped carrier typically has higher kinetic energy. If the energy lowering from the electronic potential well exceeds the kinetic energy increase, the localized state can be stabilized, leading to a polaronic state inside the gap. This localized state shares the same characteristics of a polaron but forms purely by electron-electron interactions, thus it is dubbed as \emph{electronic polaron}. Even lattice distortions still occur as a response to this electronic localization, it is simply to further stabilize the polaron by releasing residual forces.   

\begin{figure}[tbp]
\includegraphics[width=1.0\columnwidth]{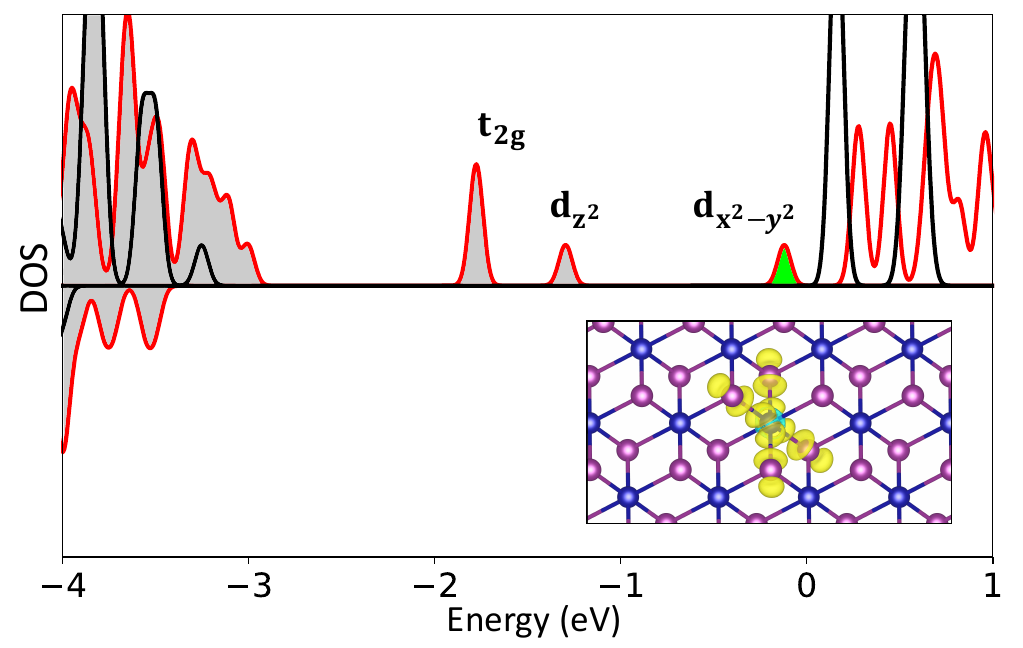}
\caption{\label{fig:fig2} (Color online) The electronic density of state (DOS) calculated by HSE06 for undoped (black) and doped (red) CrI$_2$ monolayer with atomic positions fixed. For the doped case, the valence states and polaronic state are shadowed by gray and green, respectively. Inset shows the charge distribution of the polaronic state. Atomic relaxation leads to substantial energy lowering of the in-gap states, but has little effect on their localized nature.}
\end{figure}

Such increased electron-electron interactions are expected to be pronounced in strongly-correlated systems containing active (partially occupied) $d$-shells and in systems with lower dimensionality, with reduced screening of electrons. As such, two-dimensional transition metal halides, such as CrI$_2$, CoCl$_2$, and CoBr$_2$, offer an excellent platform to realize electronic polarons. Here we focus on Crl$_2$ monolayer which is in the 1-$T$ phase with Jahn-Teller distortions. The doubly-degenerate $e_g$ orbitals of Cr split into one $d_{z^2}$ (as part of valence band) and one $d_{x^2-y^2}$ (as part of conduction band) with the in-plane inter-axial angle reduced from $60^{\circ}$ to $55^{\circ}$. Figure. 2 shows the calculated DOS of undoped (black) and doped (red) CrI$_2$ monolayer with atomic positions fixed. Without doping, CrI$_2$ is an insulator with 3.4 eV band gap. The lowest conduction band is dominated by the empty $d_{x^2-y^2}$ orbital, while the valence bands have significant hybridization between $d_{z^2}$ and $p$ orbitals. When doped, the extra electron occupies the $d_{x^2-y^2}$ state and then, due to the response of valence electrons, drops into the polaronic state inside the gap. We find three localized in-gap states dominated by $t_{2g}$, $d_{z^2}$ and $d_{x^2-y^2}$ orbitals, respectively. However, only the $d_{x^2-y^2}$ state is considered as polaron since it comes down from the conduction band, while former two are valence states popped up from the valence bands due to the strong Coulomb repulsion of $d$ orbitals. The localized nature of polaronic state is clearly shown by the charge distribution, even the atomic relaxations are not included. 

\begin{table}
	\centering
	\caption{Spontaneous localization energy, $E_L$, and atomic relaxation energy, $E_R$, of the electronic polaron. Calculations are performed with HSE06 and DFT+U. All the values are given in unit of eV.}
	\renewcommand\arraystretch{1.0}
	\begin{ruledtabular}
		\begin{tabular}{lccccccccccccccccccccccccccc}
			&System &Functional &$E_L$ &$E_R$ \\
			\hline			
			&CrI$_2$  &HSE06 &-0.15 &-0.36 \\
			&         &DFT+U &-0.92 &-0.25 \\
			\hline
			&CoCl$_2$ &HSE06 &-0.03 &-0.22 \\
			&         &DFT+U &-0.42 &-0.26 \\
			\hline
			&CoBr$_2$ &HSE06 &-0.16 &-0.16 \\
			&         &DFT+U &-0.54 &-0.19 \\
		\end{tabular}
	\end{ruledtabular}
\end{table}

Table \uppercase\expandafter{\romannumeral1} shows the spontaneous electronic localization energy ($E_{L}$), defined as the energy difference between the localized and delocalized states, and the atomic relaxation energy ($E_R$), defined as the energy difference between a fully relaxed and a structurally-unrelaxed electronic polaron. The $E_L$ and $E_R$ of CrI$_2$ are calculated to be $-0.15$ and $-0.36$ eV. To understand the role played by electronic and phononic response, we develop a method to decompose the potential well into different contributions. First, we decompose the total charge density of a supercell with a single polaron doping, $\rho_{total}$, into three parts:
\begin{equation}
	\rho_{total}= \rho_{bulk} +\rho_{pol} +\rho_{res},
\end{equation}
where $\rho_{bulk}$ is the total charge density of undoped supercell, $\rho_{pol}$ is the charge density of polaron, and $\rho_{res}$ is the change in bulk charge density due to the polaron. Here the $\rho_{res}$ doesn't contain any net charge, but only the dipoles due to relaxation. The electrostatic potential $V_{res}$ associated with $\rho_{res}$ can be obtained by solving Poisson's equation
\begin{equation}
\nabla^{2} V_{res} = -\rho_{res} / \epsilon_{0}.
\end{equation}
If $V_{res}$ is sufficiently deep, this well can overcome the kinetic energy increase associated with localization, resulting in a polaron. By decomposing $\rho_{res}$ and $V_{res}$ into electronic and phononic components, we can clearly see their relative contributions to the localization of polaron, see Fig. 3(a).

\begin{figure}[tbp]
\includegraphics[width=1.0\columnwidth]{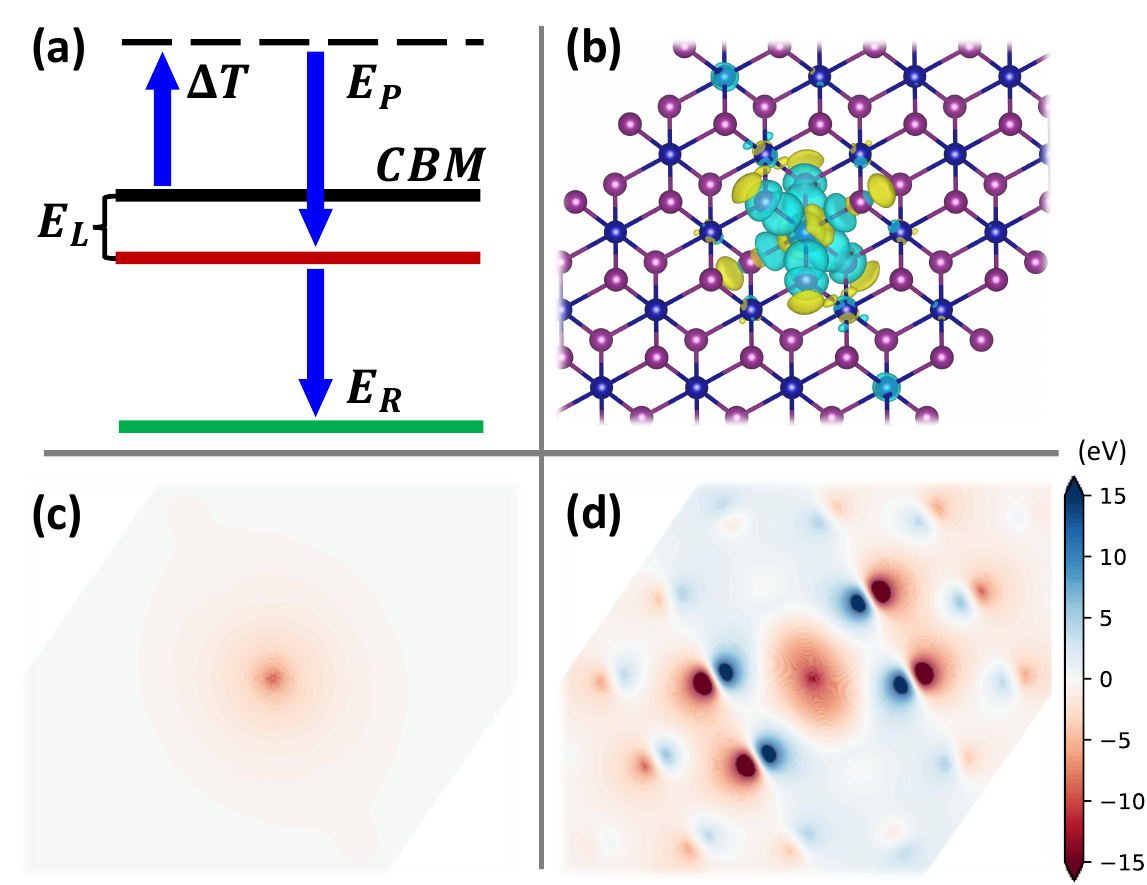}
\caption{\label{fig:fig3} (Color online) (a) The relationship of different energy states related to the formation of electronic polaron. The kinetic energy increase is denoted as $\Delta T$, the electronic and phononic potential energy lowering are given by $E_P$ and $E_R$. The red line is the electronic polaron with fixed atomic position, the green line is the fully-relaxed polaron with lattice distortion. (b) The electronic component of $\rho_{res}$ in CrI$_2$ monolayer with fixed atomic positions. (c) and (d) show the $V_{res}$ calculated without and with the lattice distortion, respectively.}
\end{figure}

With atomic positions fixed, $\rho_{res}$ does not include the effect of lattice distortions but only the contribution of electronic response, for which the result is shown in Fig. 3(b) with yellow (blue) part denoting the space area of electron accumulation (reduction). Due to the localized nature of polaron, $\rho_{res}$ is confined to a small region around the polaron. As predicted, the bulk electrons are repulsed out by the polaron, leading to dipoles towards the center. Since the polaron primarily occupies the Cr 3$d$ orbital, $V_{res}$ within the Cr sublayer, shown in Fig. 3(c), has the largest effect on polaron energy. We find that, without atomic relaxation, the electronic potential well is $-12$ eV and decays to zero away from the polaron center. To confirm that the electron-electron interaction has a major contribution to polaron formation, the $V_{res}$ with atomic relaxation is given in Fig. 3(d) which now becomes $-17$ eV. The excess potential dropping due to lattice distortion is only $-5$ eV, revealing that the electronic response dominates the potential energy lowering. Note that the dark four-fold features in Fig. 3(d) are response to atomic displacements and as they are away from the central Cr, they have little effect on the polaron energy.  

The results of $E_{L}$ and $E_{R}$ in Table \uppercase\expandafter{\romannumeral1} can be understood by further separating the $E_L$ term into two parts as
\begin{equation}
	E_{L}= \Delta T + E_{P},
\end{equation}
where $\Delta T$ denotes the kinetic energy increase caused by electronic localization, and $E_{P}$ denotes the energy lowering by the electronic potential well. As sketched in Fig. 3(a), there is a large cancellation between positive $\Delta T$ and negative $E_P$ resulting in the relatively small $E_L$, which represents the stability of electronic polaron. It is important to note, however, $E_L$ should not be directly compared with $E_R$. Physically $E_P$ and $E_R$ are directly comparable quantities corresponding to potential energy lowering by electronic and phononic response, respectively. Determination of $E_P$ requires the knowledge of $\Delta T$, which can be difficult to accurately calculate. To approximate $\Delta T$, we turn to the tight-binding model, where the kinetic energy increase by localization equals half of the band width. From the HSE06 band structure [see Fig. S1 in Supplemental Material\cite{ourSM}], the lowest conduction band originates from Cr $3d_{x^{2}-y^{2}}$ orbital and well separates from the higher energy $4s$ bands. The calculated band width is $0.58$ eV corresponding to $\Delta T \approx 0.29$ eV. Therefore, $E_P$ and $E_R$ are determined as $-0.44$ and $-0.36$ eV for CrI$_2$, roughly in line with the results of $V_{res}$ in Fig. 3(c) and (d), confirming that electronic response is the major mechanism of polaron formation. The negative $E_L$ clearly indicates that $E_P$ can solely stabilize the polaron without the help of lattice distortions.

\begin{figure}[tbp]
\includegraphics[width=1.00\columnwidth]{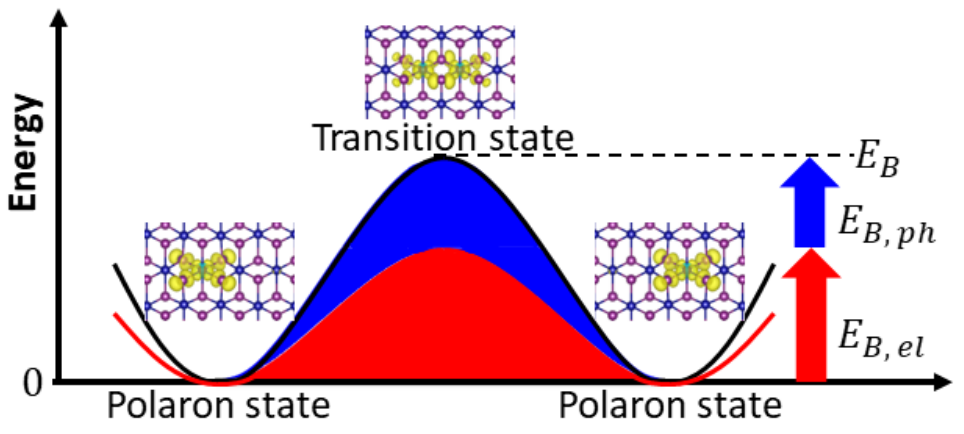}
\caption{\label{fig:fig4} (Color online) A schematic of the inter-site hopping of electronic polaron. The polaron state denotes an electronic polaron locating on one Cr atom or on its nearest neighborhood. The transition state corresponds the case of electron occupying the two Cr sites equally. $E_B$ is the adiabatic barrier which is separated into electronic (red) and phononic (blue) contributions.} 
\end{figure}

Dynamic characteristic is also an important aspect of electronic polaron, since it determines to what an extent the polaronic state can be viewed as localized in experiments. The adiabatic barrier of the inter-site hopping, defined as the energy difference between electronic polaron and transition state\cite{franchini2021polarons}, is shown in Fig. 4. The calculated barrier height is $E_{B}=227$ meV, on the same order of another prototypical material TiO$_2$\cite{deskins2007electron}. Considering the carrier mobility $\mu$ is sensitive to temperature $T$ that $\mu \propto e^{-\frac{E_{B}}{kT}}$ and a low temperature is applied in our experiments, such barrier should provide a long enough life-time for polarons. This barrier can be further separated into the electronic $E_{B,el}$ and phononic contribution $E_{B,ph}$ as $E_{B}=E_{B,el}+E_{B,ph}$. By calculating the energies of the polaron localized at a lattice site and at the transition state, including and excluding phonon relaxation, we determine the contributions to the adiabatic barrier to be 125 meV and 102 meV for the electronic and phononic terms, respectively. Thus, combined with the energetic results above, we confirm that electronic response has a major contribution to the formation of electronc polaron both energetically and dynamically.

Results are also given for Co-based systems CoCl$_2$ and CoBr$_2$, which show similar characteristic as CrI$_2$. In both cases, the negative $E_L$ implies the spontaneous localization of doped electron without the involvement of phonon. In addition, the $E_L$ and $E_R$ are also calculated at the DFT+U level. We find that calculated $E_L$ are also negative. These results demonstrate the existence of electronic polaron, which is insensitive to the functional used in the calculation. Finally, the two-dimensional nature of this family of materials may also contribute to the formation of polarons, as it has been shown that for spatial dimension $N \leq 2$, that any attractive local potential leads to the existence of a localized bound state\cite{chadan2003}.  

\begin{figure}[tbp]
\includegraphics[width=0.90\columnwidth]{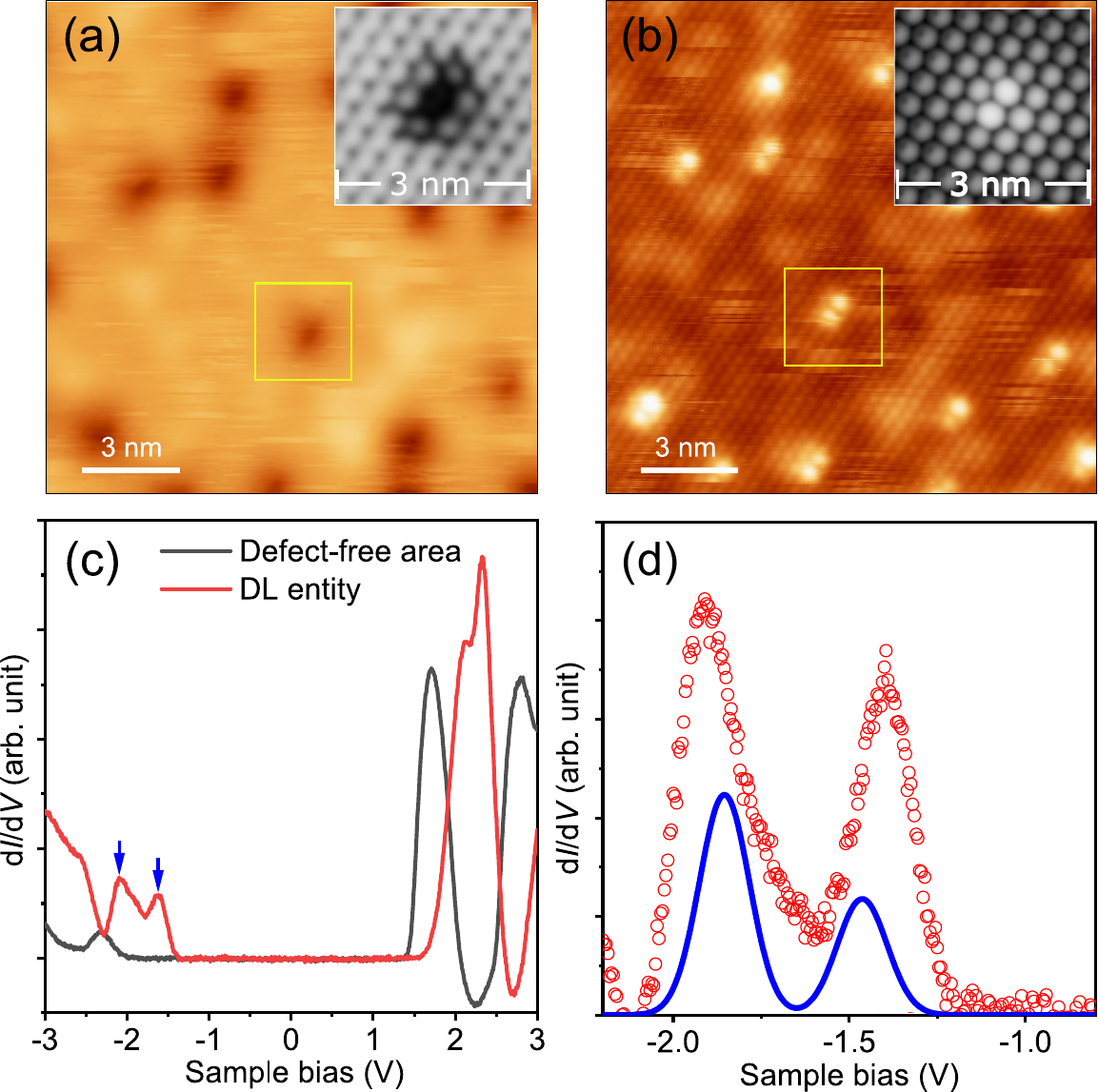}
\caption{\label{fig:fig5} (Color online) (a, b) The STM topography of DL entities in CrI$_2$ single layer at positive (a) (V$_s$ = 1.6 V, I$_t$ = 50 pA) and negative (b) (V$_s$ = $-$3.0 V, I$_t$ = 100 pA) sample bias. An isolated DL entity is denoted by yellow square. Inset shows the simulated STM topography of polaron. (c) Differential conductance spectra (set point condition: V$_s$ = 0.8 V, I$_t$ = 3 pA) measured on defect-free area (black) and a typical DL entity (red). The blue arrows mark two in-gap states. (d) Comparison between experiment and theory for CrI$_2$. Red circles and bule curve denote, respectively, the measured differential conductance spectrum of DL entity and calculated $dI/dV$ spectrum of polaron. Due to the large $9 \times 9$ supercell required, the calculations here are performed with DFT+U method.} 
\end{figure}

Recently, we have achieved the growth of high-quality CrI$_2$ single layer on graphene covered SiC(0001) substrate via molecular beam epitaxy. Here we examine the validity of our first-principles calculations via STM/STS measurements, for which the technical details have been given in a previous publication\cite{peng2020mott}. From the STM topography in Figs. 5(a) and (b), CrI$_2$ single layer shows a considerable density of intrinsic defect like (DL) entities. These DL entities appear as depression spots at a positive bias of 1.6 V, corresponding to the energy of conduction bands, and protrusion dimers at a negative bias of $-$3.0 V, corresponding to the energy of valence bands. Differential conductance spectrum of defect-free area has an insulating gap of $\sim$3.3 eV, while the spectrum of a typical DL entity indicates two prominent in-gap states separated by $\sim$0.5 eV [Fig. 5(c)]. In Fig. 5(d), we compare the calculated $dI/dV$ spectrum of polaron with the measured differential conductance spectrum of a typical DL entity. Here the polaron is calculated with fully relaxed lattice structure, i.e., both electron-electron correlation and electron-phonon coupling are included. The features of in-gap states are reasonably captured by our calculations.  Experimentally, upon approaching a DL entity, the band of CrI$_2$ rigidly shifts towards higher energy [Fig. S2], suggesting the presence of a trapped electron inside the DL entity. Such rigid band shift in proximity to polaron is also reproduced with our calculations [Fig. S3]. The insets of Figs. 5(a) and (b) show the calculated STM topography of polaron at positive and negative bias, respectively. The polaron displays depressions at positive bias and dimer protrusions at negative bias, which agrees with the experimental results. Furthermore, experimentally all the dimers have the same orientation, which is consistent with our calculations where the polaron charge is dimer-like and perpendicular to the long-bond direction [Fig. S4]. These agreements are consistent with the in-gap states being of polaronic nature.

\begin{figure}[tbp]
	\includegraphics[width=1.0\columnwidth]{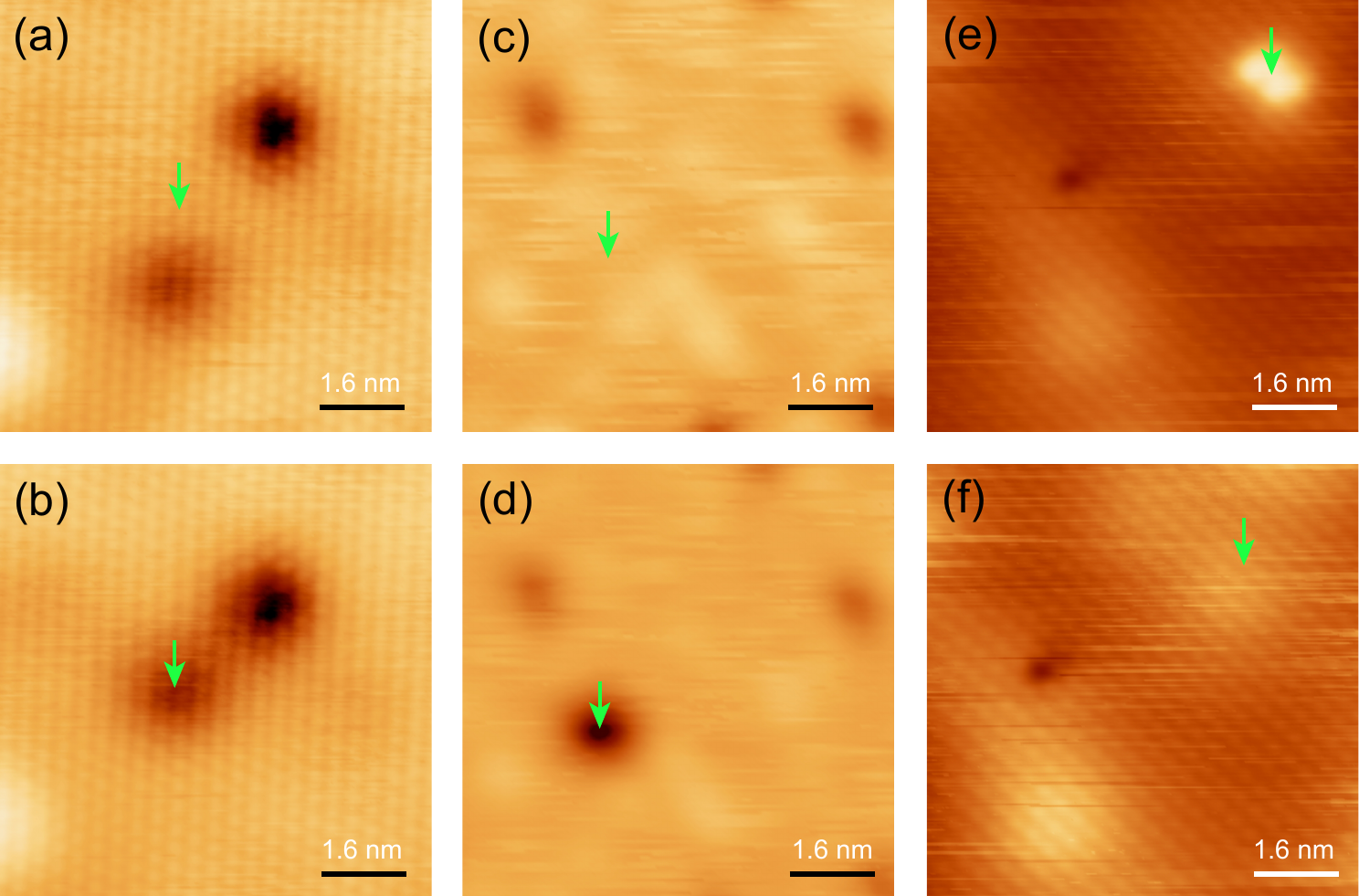}
	\caption{\label{fig:fig6} (Color online) (a, b) STM images (V$_s$ = 1.75 V, I$_t$ = 100 pA) showing prior to (a) and after (b) moving a DL entity with tip. (c, d) STM images (V$_s$ = 1.6 V, I$_t$ = 50 pA) showing prior to (c) and after (d) generating a DL entity with tip. (e, f) STM images (V$_s$ = $-3$ V, I$_t$ = 50 pA) showing prior to (e) and after (f) erasing a DL entity with tip. Note that the DL entity appears as a protrusion at $-3$ V. For all images, the tip is placed at the positions indicated with green arrows during the manipulations.}
\end{figure}

The identification of the in-gap states as polaronic in nature is further bolstered by our ability to directly manipulate them with the STM tip. Indeed, a DL entity can be moved by the tip when we locate the tip close to it and gradually increase the sample bias voltage to above 3.5 V [Figs. 6(a) and (b)]. We can also controllably create a DL entity with the tip by gradually increasing the sample bias voltage to above 3.5 V at a defect-free area [Figs. 6(c) and (d)]. Moreover, a DL entity can be eliminated by positioning the tip above it and adding a large enough negative bias voltage $\sim -3.5$ V [Figs. 6(e) and (f)]. This ability to revesibly create and eliminate DL entities strongly suggests that the DL entity is not associated with nonstoichiometry, but instead has an electronic origin possibly coupled with a local lattice distortion. These observations, along with the agreement of the simulated STM/STS of the electronic polaron with experiments, provide strong evidence that the observed DL entity is indeed the predicted electronic polaron.

In summary, we propose a new form of polaron, an electronic polaron, in strongly correlated materials. Instead of the carrier being trapped by a lattice distortion, for the electronic polaron the carrier is spontaneously localized due to electronic polarization. By first-principles calculations combined with STM/STS experiments, we present evidence for their existence in transition metal halides, particularly in CrI$_2$ monolayer. The identification of defect-like features in CrI$_2$ as electron polarons in nature is further bolstered by the ability to manipulate and reversibly create/annihilate them via the STM tip. Furthermore, as these polarons are relatively immobile, the ability to create, manipulate, and destroy spin-polarized charged point defects on demand with atomic precision may have profound impacts for future electronic devices.

\begin{acknowledgments}
This material is based upon theoretical work supported by the U.S. Department of Energy, Office of Science, Office of Basic Energy Sciences under Award Number DE-SC-0002623. The supercomputer time sponsored by the National Energy Research Scientific Center (NERSC) under DOE Contract No. DE-AC02-05CH11231 and the Center for Computational Innovations (CCI) at RPI are also acknowledged. Experimental work in China was funded by the National Key Research and Development Program of China (Grant Nos. 2017YFA0403501 and 2016YFA0401003), the National Science Foundation of China (Grant Nos. 11874161 and U20A6002). We thank Wenhao Zhang, Yuan Yuan, Zhenyu Liu and Huinan Xia for technical assistance in experiments. 

Y. L. and M. C. contributed equally to this work. 
\end{acknowledgments}


\begin{thebibliography}{90}%
\makeatletter
\bibitem{emin2013polarons} D. Emin, \emph{Polarons} (Cambridge University Press, UK, 2013).

\bibitem{giustino2017electron} F. Giustino, \emph{Electron-phonon interactions from first principles}, Rev. Mod. Phys. \textbf{89}, 015003 (2017).

\bibitem{morgan2009small} B. J. Morgan, D. O. Scanlon, and G. W. Watson, \emph{Small polarons in Nb- and Ta-doped rutile and anatase TiO$_2$}, J. Mater. Chem. \textbf{19}, 5175 (2009).

\bibitem{lee2006doping} P. A. Lee, N. Nagaosa, and X.-G. Wen, \emph{Doping a Mott insulator: Physics of high-temperature superconductivity}, Rev. Mod. Phys. \textbf{78}, 17 (2006).

\bibitem{kravchenko2017strongly} S. Kravchenko, \emph{Strongly correlated electrons in two dimensions} (CRC Press, New York, 2017).

\bibitem{mao2020interfacial} Y. Mao, X. Ma, D. Wu, C. Lin, H. Shan, X. Wu, J. Zhao, A. Zhao, and B. Wang, \emph{Interfacial polarons in van der waals heterojunction of monolayer SnSe$_2$ on SrTiO$_3$ (001)}, Nano Lett. \textbf{20}, 8067 (2020).

\bibitem{kang2018holstein} M. Kang, S. W. Jung, W. J. Shin, Y. Sohn, S. H. Ryu, T. K. Kim, M. Hoesch, and K. S. Kim, \emph{Holstein polaron in a valley-degenerate two-dimensional semiconductor}, Nat. Mater. \textbf{17}, 676 (2018).

\bibitem{Frohlich1} H. Fröhlich, H. Pelzer, and S. Zienau, \emph{XX. Properties of slow electrons in polar materials}, The London, Edinburgh, and Dublin Philosophical Magazine and Journal of Science \textbf{41}, 221 (1950).

\bibitem{Frohlich2} H. Fröhlich, \emph{Electrons in lattice fields}, Advances in Physics \textbf{3}, 325 (1954).

\bibitem{Holstein1} T. Holstein, \emph{Studies of polaron motion: Part I. The molecular-crystal model}, Annals of Physics \textbf{8}, 325 (1959).

\bibitem{Holstein2} T. Holstein, \emph{Studies of polaron motion: Part II. The ``small'' polaron}, Annals of Physics \textbf{8}, 343 (1959).

\bibitem{resta2007theory} R. Resta and D. Vanderbilt, \emph{Theory of polarization: A modern approach}, in Physics of Ferroelectrics (Springer, 2007).

\bibitem{berciu2009electronic} M. Berciu, I. Elfimov, and G. A. Sawatzky, \emph{Electronic polarons and bipolarons in iron-based superconductors: the role of anions}, Physical Review B \textbf{79}, 214507 (2009).

\bibitem{kresse1996efficient} G. Kresse and J. Furthmüller, \emph{Efficient iterative schemes for ab initio total-energy calculations using a plane-wave basis set}, Phys. Rev. B. \text{54}, 11169 (1996).

\bibitem{heyd2003hybrid} J. Heyd, G. E. Scuseria, and M. Ernzerhof, \emph{Hybrid functionals based on a screened coulomb potential}, J. Chem. Phys. \textbf{118}, 8207 (2003).

\bibitem{heyd2006erratum} J. Heyd, G. E. Scuseria, and M. Ernzerhof, \emph{Erratum: ``hybrid functionals based on a screened coulomb potentia''[j. chem. phys. 118, 8207 (2003)]}, J. Chem. Phys. \textbf{124}, 219906 (2006).

\bibitem{tran2006hybrid} F. Tran, P. Blaha, K. Schwarz, and P. Novák, \emph{Hybrid exchange-correlation energy functionals for strongly correlated electrons: Applications to transition-metal monoxides}, Phys. Rev. B \textbf{74}, 155108 (2006).

\bibitem{bilc2008hybrid} D. I. Bilc, R. Orlando, R. Shaltaf, G. M. Rignanese, J. Íñiguez, and P. Ghosez, \emph{Hybrid exchange-correlation functional for accurate prediction of the electronic and structural properties of ferroelectric oxides}, Phys. Rev. B \textbf{77}, 165107 (2008).

\bibitem{zhuang2016strong} H. L. L. Zhuang, P. Kent, and R. G. Hennig, \emph{Strong anisotropy and magnetostriction in the two-dimensional stoner ferromagnet Fe$_3$GeTe$_2$}, Phys. Rev. B \textbf{93}, 134407 (2016).

\bibitem{hellgren2017critical} M. Hellgren, J. Baima, R. Bianco, M. Calandra, F. Mauri, and L. Wirtz, \emph{Critical role of the exchange interaction for the electronic structure and charge-density wave formation in TiSe$_2$}, Phys. Rev. Lett. \textbf{119}, 176401 (2017).

\bibitem{dudarev1998electron} S. L. Dudarev, G. A. Botton, S. Y. Savrasov, C. J. Humphreys, and A. P. Sutton, \emph{Electron-energy-loss spectra and the structural stability of nickel oxide: An LSDA+U study}, Phys. Rev. B \textbf{57}, 1505 (1998).

\bibitem{ourSM} See Supplemental Material at **URL**.

\bibitem{franchini2021polarons} C. Franchini, M. Reticcioli, M. Setvin, and U. Diebold, \emph{Polarons in materials}, Nature Reviews Materials \textbf{6}, 560 (2021).

\bibitem{deskins2007electron} N. A. Deskins and M. Dupuis, \emph{Electron transport via polaron hopping in bulk TiO$_2$: A density functional theory characterization}, Phys. Rev. B \textbf{75}, 195212 (2007).

\bibitem{chadan2003} K. Chadan, N. N. Khuri, A. Martin, and T. Tsun Wu, \emph{Bound states in one and two spatial dimensions}, J. Math. Phys. \textbf{44}, 406 (2003).

\bibitem{peng2020mott} L. Peng, J. Z. Zhao, M. Cai, G. Y. Hua, Z. Y. Liu, H. N. Xia, Y. Yuan, W. H. Zhang, G. Xu, L. X. Zhao, et al., \emph{Mott phase in a van der waals transition-metal halide at single-layer limit}, Phys. Rev. Research \textbf{2}, 023264 (2020)
\end{thebibliography}
\end{document}